\documentclass[11pt]{amsart}
\usepackage{amssymb,latexsym,amsmath,amsthm,mathrsfs,yfonts}
\usepackage{fullpage,enumerate}
\input xy
\xyoption{all}

\newcommand{\beq}{\begin{eqnarray}}
\newcommand{\eeq}{\end{eqnarray}}

\newcommand\om{\omega}
\newcommand\Om{{\it \Omega}}
\newcommand\two{\text{I}\!\text{I}}
\newcommand\id{\mathrm{id}}
\def\colim{{\varinjlim}}

\def\H{\textup{H}}
\def\RK{\textup{RK}}
\def\sC{\text{$\sigma$-$C^*$}}
\def\K{\textup{K}}
\def\cpt{{\mathcal K}}
\def\CT{\textup{CT}}

\newcommand\RR{\mathbb R}
\newcommand\CC{\mathbb C}
\newcommand\ZZ{\mathbb Z}
\newcommand\TT{\mathbb T}

\begin{document}

\title{Topology and Flux of T-Dual Manifolds with Circle Actions}

\author{Varghese Mathai}
\address[V Mathai]{Department of Pure Mathematics, University of Adelaide,
Adelaide, SA 5005, Australia}
\email{mathai.varghese@adelaide.edu.au}

\author{Siye Wu}
\address[S Wu]{Department of Mathematics, University of Hong Kong,
Pokfulam Road, Hong Kong, China}
\email{swu@maths.hku.hk}

\begin{abstract}
We present an explicit formula for the topology and H-flux of the T-dual of
a general type \two\, compactification, significantly generalizing earlier
results.
Our results apply to T-dualities with respect to {\em any} circle action
on spacetime $X.$ 
As before, T-duality exchanges type \two A and type \two B string theories.  
A new consequence is that the T-dual spacetime is a singular space when the
fixed point set $X^\TT$ is non-empty; the singularities correspond to Kaluza-Klein monopoles.
We propose that the Ramond-Ramond charges of type \two\ string theories on
the singular dual are classified by twisted equivariant cohomology groups.
We also discuss the K-theory approach.
\end{abstract}

\thanks{{\em Acknowledgements.}
This research was supported under Australian Research Council's Discovery
Projects funding scheme (project number DP0878184).
V.M.\ is the recipient of an Australian Research Council Australian
Professorial Fellowship (project number DP0770927).
S.W.\ is supported in part by a General Research Fund from the Research
Grants Council of Hong Kong (project number HKU706010P).
The authors would like to thank Jarah Evslin and Jonathan Rosenberg for
useful feedback.}

\maketitle

It is by now well known that there are five (super) string theories related
to each other by transformations called dualities. 
Two string theories are said to be dual to each other if they are equivalent
descriptions of the same theory at the quantum level.

The global aspects of one such duality, called T-duality in type \two\,
string theories, involves compactifications of spacetime $X$ with an H-flux. 
It is a generalization of the $R\to 1/R$ transformation of string theories
compactified on a circle of radius $R$.
The local transformation rules of the low energy effective fields under
T-duality, known as the Buscher rules \cite{Bus}, have been known for some
time.
However, in cases in which there is a topologically nontrivial NS 3-form
H-flux, the Buscher rules only make sense on each local spacetime patch.
Several examples of T-duals to such backgrounds have been found  
\cite{AABL,DLP,DRS,GLMW,KSTT} and in each case it was seen that T-duality  
changes not only the H-flux but also the spacetime topology.  
In \cite{BEM}, a general formula was presented for the topology and
H-flux of a compactification from the topology and H-flux of its T-dual,
with respect to any free circle action, in particular reproducing the
examples in the literature mentioned above.  
As supporting evidence, it was shown there that locally, the said formula
agrees with the Buscher rules and that globally it yields an isomorphism
of the twisted cohomology and twisted K-theory-valued conserved Ramond-Ramond
charges \cite{MM,Wib,BM,BS,Pan}.

In this paper, we will produce an explicit formula for the topology and
H-flux of the T-dual of a more general type \two\, compactification,
significantly generalizing earlier results. 
Our results apply to T-dualities with respect to {\em any} circle action.
We start with the spacetime $X$, equipped with a flux form $H$, that is a
manifold without boundary such that the fixed point set $X^\TT$ is non-empty.
The T-dual spacetime $\widehat X/\TT$ is a stratified space, possibly with
boundary, in the sense of \cite{GM}, with a dual flux $\widehat H$ which is
an equivariantly closed differential form on $\widehat X$.
The singularities of the spacetime and that of the flux form correspond
to Kaluza-Klein monopoles \cite{Sorkin,Pan}. 
Topological aspects of examples from mirror symmetry (when $X$ is Calabi-Yau,
cf.~\cite{Hori}) fit into our framework, solving an open problem stated
in \cite{BEM}.
Technically, the new approach in this paper is to use the correspondence
space in a fundamental way in our construction of the T-dual spacetime,
in contrast with the construction of the T-dual in \cite{BEM}.
We show that the twisted cohomology of $(X,H)$ is isomorphic, with a degree
shift, to the twisted equivariant cohomology of $(\widehat X,\widehat X)$,
and a similar result in twisted equivariant K-theory.
We thus propose that {\em the Ramond-Ramond charges of type \two\ strings
theories on the singular space $\widehat X/\TT$, with the flux form
$\widehat H$, are classified by the twisted equivariant cohomology groups.}
Possible relevance of equivariant cohomology in the context of T-duality
was first mentioned in \S2, \cite{BHM}.
Intersection cohomology \cite{GM} is another good replacement for the
cohomology of $\widehat X/\TT$, and it is interesting to see if it is possible
to realize T-duality isomorphisms in terms of intersection cohomology and its
twisted variants.

Consider a pair $(X,H)$, where $X$ is a spacetime and $H$ is a background
flux, a closed differential $3$-form on $X$ with integral periods.
We also suppose that there is a smooth action of a circle $\TT$ on $X$
preserving $H$ (by averaging over $\TT$, this can always be arranged
without changing the cohomology class of $H$). 
We now construct the T-dual of $(X,H)$ in such a way that it reduces to the
case studied in \cite{BEM} whenever the $\TT$-action is free.
Consider the product $X \times E\TT$ with the diagonal $\TT$-action, which
is free since the action on the universal space $E\TT$ is free.
The universal $\TT$-space $E\TT$ can be chosen to be either the unit sphere in 
an infinite dimensional Hilbert space, or the direct limit of odd dimensional 
spheres $S^\infty =\colim\,S^{2n+1}$.
Therefore $X\times E\TT$ is a smooth countably compactly generated manifold
and $p_1^*(H)$, where $p_1\colon X\times E\TT\to X$ is the projection, is
a background flux on it.
The quotient space
$$     X_\TT =\left(X\times E\TT\right)/\TT,     $$
where $\TT$ acts diagonally on $X\times E\TT$, is called the Borel construction.
Note that there is a fibration
$$     X \hookrightarrow X_\TT \longrightarrow B\TT,      $$
where the classifying space $B\TT$ can be chosen to be either the projective
Hilbert space, or the direct limit of complex projective spaces 
$\CC P^\infty=\colim\,\CC P^n$.
The equivariant cohomology is by fiat
$\H^\bullet_\TT(X,\ZZ)=\H^\bullet(X_\TT,\ZZ)$.
It has the following properties, cf.~\cite{AB}.
\begin{enumerate}
\item
The equivariant cohomology $\H^\bullet_\TT(X,\ZZ)$ is a module over 
$\H^\bullet(B\TT,\ZZ)=\ZZ[\phi]$.
\item
The inclusion map $i\colon X\hookrightarrow X_\TT$ induces the restriction
$i^*\colon\H^\bullet_\TT(X,\ZZ)\to\H^\bullet(X,\ZZ)$.
\item
The surjective map $\sigma\colon X_\TT\to X/\TT$ induces a map
$\sigma^*\colon\H^\bullet(X/\TT,\ZZ)\to\H^\bullet_\TT(X,\ZZ)$,
which is an isomorphism whenever the $\TT$-action on $X$ is free.
\item
Upon localisation, there is an isomorphism
$\H^\bullet_\TT(X)[\phi^{-1}]\cong\H^\bullet(X^\TT)[\phi,\phi^{-1}]$,
where $X^\TT$ denotes the fixed point subset of $X$.
\end{enumerate}
These properties explain why the equivariant cohomology $\H_\TT(X)$ is a good
replacement of the cohomology of $X/\TT$ when the latter is singular.

The {\em Gysin sequence} of the $\TT$-fibration $X\times E\TT\to X_\TT$ is
the long exact sequence
$$ \to\H^j(X)\stackrel{\pi_*}\to\H^{j-1}(X_\TT)\stackrel{\cup\,e_\TT}
   \to\H^{j+1}(X_\TT)\stackrel{\pi^*}\to\H^{j+1}(X)\to                 $$
where $\cup\, e_\TT$ denotes the cup product with the (equivariant) Euler
class $e_\TT\in\H^2(X_\TT,\ZZ)=\H^2_\TT(X,\ZZ)$.
In this case, $e_\TT=\phi$, which replaces the curvature of the (singular)
$\TT$-fibration $X\to X/\TT$.
So the Gysin sequence becomes
$$ \to\H^j(X)\stackrel{\pi_*}\to\H^{j-1}_\TT(X)\stackrel{\cup\,\phi}
   \to\H^{j+1}_\TT(X)\stackrel{\pi^*}\to\H^{j+1}(X)\to.                $$

We next construct the T-dual of $(X,H)$.
We recall the classification of equivariant circle bundles on a
$\TT$-manifold $X$.

\medskip

\noindent{\bf Proposition 1 (Equivariant circle bundles)}
{\em Let $X$ be a connected $\TT$-manifold.
The equivariant first Chern class $c_1^\TT$ gives rise to a one-to-one
correspondence between equivalence classes of $\TT$-equivariant circle
bundles over $X$ and elements of $\H_\TT^2(X,\ZZ)$.}

\medskip

For locally finite CW-complexes with continuous $\TT$-actions, this result
was proved in \cite{HY}.
For smooth actions of connected groups, Proposition~1 is due to \cite{Ri}.
There are different proofs, due to \cite{GGK}, Theorem C.47, and \cite{HL}.
None of these proofs are particularly simple.

By Proposition~1, there is a $\TT$-equivariant circle bundle $\widehat X$ over
$X$ such that $c_1^\TT(\widehat X)=\pi_*([H])\in\H^2_\TT(X,\ZZ)$.
The next goal is to define a class $[\widehat H]\in\H^3_\TT(\widehat X,\ZZ)$
corresponding to a flux on the singular space $\widehat X/\TT$.
Consider the commutative diagram
\begin{equation*}
\qquad\xymatrix @=4pc @ur { X \ar[d] & 
 \widehat X \ar[d] \ar[l]_p \\ X/\TT & \widehat X/\TT \ar[l]}
\end{equation*}
containing the singular spaces $X/\TT$ and $\widehat X/\TT$.
We can replace the singular spaces by their Borel constructions by lifting the
previous diagram to  
\begin{equation}\label{correspondenceb}
\qquad\xymatrix @=4pc @ur { X \times E\TT \ar[d]_\pi & 
\widehat X\times E\TT\ar[d]^{\widehat p} \ar[l]_{p\times\id} \\
X_\TT & \widehat X_\TT \ar[l]^{\widehat \pi}}.
\end{equation}
The conditions that uniquely specify $[\widehat H]$ are:
\begin{enumerate}
\item $\widehat\pi_*([\widehat H])=e_\TT\in\H^2_\TT(X,\ZZ)$;
\item $(p\times\id)^*([H])=i^*([\widehat H])\in\H^3(\widehat X,\ZZ)$,
where $i\colon\widehat X\to\widehat X_\TT$ is the inclusion map.
That is, $[\widehat H]$ is an equivariant extension of $p^*([H])$ on 
$\widehat X$.
\end{enumerate}

Recall that the Cartan model for equivariant cohomology (with real
coefficients) is 
$$ \H_\TT^\bullet(X)=\H^\bullet(\Om^{\bullet}(X)^\TT[\phi],d-\phi\,\iota_V), $$
where $\Om^{\bullet}(X)^\TT[\phi]$ denotes polynomials in $\phi$ with
coefficients that are $\TT$-invariant differential forms on $X$ and $V$ is
the vector field on $X$ generating the $\TT$-action.
The class $c_1^\TT(\widehat X)=\pi_*([H])$ is represented by $\iota_VH$.
On $\widehat X\to X$, there is a $\TT$-invariant connection
$\widehat A\in\Om^1(\widehat X)^\TT$ whose (equivariant) curvature is
$\widehat F=d\widehat A=\iota_VH$, with the property $\iota_V\widehat A=0$.
We can choose an equivariantly closed $3$-form
$\widehat H=\widehat A\wedge\phi+H\in\Om^3(\widehat X)^\TT[\phi]$
representing $[\widehat H]$.
Just as the twisted cohomology $\H^\bullet(X,H)$ is the cohomology
of the de Rham complex $\Om^\bullet(X)$ with a twisted differential
$d_H=d+H\wedge\cdot\,$, the twisted equivariant cohomology
$\H^\bullet_\TT(\widehat X,\widehat H)$ is defined as the cohomology of the
Cartan complex $\Om^\bullet(\widehat X)^\TT[\phi]$ with a twisted equivariant
differential $\widehat d_{\widehat H}=d-\phi\,\iota_V+\widehat H\wedge\cdot\,$.
They are both only $\ZZ_2$-graded cohomology groups.
Our main result is the following:

\medskip

\noindent {\bf Theorem 1 (T-duality for circle actions)}
{\em Let $X$ be a connected manifold and let $H$ be a background flux, 
that is, $H$ is a closed differential $3$-form on $X$ with integral periods.
We also suppose that there is a smooth action of a circle $\TT$ on $X$
preserving $H$.
The T-dual of $(X,H)$ is the pair $(\widehat X,\widehat H)$ constructed above, where
$\widehat\pi\colon\widehat X\to X$ is a $\TT$-equivariant circle bundle over
$X$ such that $c_1^\TT(\widehat X)=\pi_*([H])$,
$\widehat\pi_*([\widehat H])=e_\TT$ and
$p^*([H])=i^*([\widehat H])\in H^3(\widehat X)$.
There is a ``T-duality isomorphism'' between the twisted cohomology
groups
$$   \H^\bullet(X,H)\cong\H^{\bullet+1}_\TT(\widehat X,\widehat H).   $$}
We note that the T-dual of the spacetime $X$ with flux $H$ for general circle
actions, is a singular manifold $\widehat X/\TT$, whose desingularization
$\widehat X$ is of dimension one higher than that of $X$. 
Although typically, the equivariant cohomology of a $\TT$-manifold $X$ with
nonempty fixed point set $X^\TT$ is infinite dimensional, the isomorphism in
Theorem~1 implies that the twisted equivariant cohomology groups
$\H^\bullet_\TT(\widehat X,\widehat H)$ are always finite dimensional.
The isomorphism in Theorem~1 is a compelling evidence that the Ramond-Ramond
charges in the type \two\, string theories on the singular space
$\widehat X/\TT$ are classified by the twisted equivariant cohomology group
$\H_\TT^\bullet(\widehat X,\widehat H)$.
Further evidence will be given by a similar isomorphism in twisted
equivariant K-theory in the Appendix.

We explain the above isomorphism in two different ways.  
First, by the main result in \cite{BEM} and using the commutative diagram
above, we have
$$
\H^\bullet(X,H)  \cong\H^\bullet(X\times E\TT,p_1^*(H))
 \cong\H^{\bullet+1}(\widehat X_\TT,\widehat H)
 \cong\H^{\bullet+1}_\TT(\widehat X,\widehat H).
$$
The isomorphism can also be obtained by Hori type formulas \cite{Hori1},
which we now establish, in the Cartan model of equivariant cohomology.
Define the maps
\begin{align*} 
S\colon\Om^\bullet(X)^\TT\to\Om^{\bullet+1}(\widehat X)^\TT[\phi],\\
T\colon\Om^\bullet(\widehat X)^\TT[\phi]\to\Om^{\bullet+1}(X)^\TT
\end{align*} 
by
\begin{align*}
S&\colon G\longmapsto-\iota_VG+\widehat A\wedge G,\\
T&\colon\om(\phi)=\om_1(\phi)+\widehat A\wedge\om_2(\phi)\longmapsto\om_2(0),
\end{align*} 
where $\om_1(\phi),\om_2(\phi)\in\Om^\bullet(X)^\TT[\phi]$.
Then one verifies that
\begin{align*} 
S\circ d_H+\widehat d_{\widehat H}\circ S=0,\\
T\circ\widehat d_{\widehat H}+d_H\circ T=0.
\end{align*} 
Moreover,
\begin{align*}
T\circ S&=\id,\\
\id-S\circ T&=R\circ\widehat d_{\widehat H}+\widehat d_{\widehat H}\circ R,
\end{align*} 
where
$$  R\colon\Om^\bullet(\widehat X)^\TT[\phi]\to
    \Om^{\bullet+1}(\widehat X)^\TT[\phi]              $$   
is defined by
$$  R\colon\om(\phi)\longmapsto\phi^{-1}[\om_2(\phi)-\om_2(0)].   $$
Therefore $S$ and $T$ are homotopy inverses of each other and the induced
maps $S_*$ and $T_*$ on the twisted cohomology groups realize the
isomorphism in Theorem~1.

If we started with the spacetime $X$ without boundary and with $H=0$, the
T-dual spacetime $\widehat X/\TT$ is a manifold with non-empty boundary
components.
(In addition, there is a noncommutative algebra associated to the T-dual.
See, for example, \cite{Pan}.)
The dual flux is $\widehat H=\phi\wedge d\widehat\theta$, where
$d\widehat\theta$ is the standard $1$-form on $\widehat\TT$. 
Near a component of the fixed-point set, $X$ looks like $\CC^n\times M$,
where $\TT$ acts on each coordinate of $\CC^n$ and $\dim M= 10-2n$.
If the weights of the $\TT$-action are $1$, the T-dual manifold
$\widehat X/\TT$ is the product of a cone on $\CC P^{n-1}$ with
$M\times\widehat\TT$ near the singularity. 
For small $n$, a neighbourhood of the singularity is illustrated in Figure 1.

\begin{figure}[htbp]
\caption{Singularity of the T-dual space of $\CC^n\times M^{10-2n}$.
(a) $n=1$; (b) $n=2$; (c) $n=3$.}
\begin{center}
\setlength{\unitlength}{.98mm}
\begin{picture}(130,145)
\put(0,130){\circle*{3}} \put(0,130){\line(1,0){50}}
\put(60,128){$\times M^8\times\widehat\TT$\qquad\qquad(a)}
\put(0,90){\circle*{3}}
\qbezier(0,90)(30,95)(50,115) \qbezier(0,90)(30,85)(50,65)
\put(60,88){$\times M^6\times\widehat\TT$\qquad\qquad(b)}
\put(0,25){\circle*{3}}
\qbezier(0,25)(30,30)(50,50) \qbezier(0,25)(30,20)(50,0)
\put(60,23){$\times M^4\times\widehat\TT$\qquad\qquad(c)}
\thicklines
\put(32,79.5){$\CC\;\!\!\mathrm{P}^1$}
\qbezier(38.2,104)(45,111.86)(46,90) \qbezier(38.2,76)(45,68.14)(46,90)
\put(35,90){\circle*{1}}
\put(35.05,92){\circle*{1}}\put(35.18,94){\circle*{1}}
\put(35.42,96){\circle*{1}}\put(35.77,98){\circle*{1}}
\put(36.27,100){\circle*{1}}\put(37,102){\circle*{1}}
\put(35.05,88){\circle*{1}}\put(35.18,86){\circle*{1}}
\put(37,78){\circle*{1}} 
\put(31,14.5){$\CC\;\!\!\mathrm{P}^2$}
\qbezier(38.2,39)(45,46.86)(46,25) \qbezier(38.2,11)(45,3.14)(46,25)
\put(35,25){\circle*{1}}
\put(35.05,27){\circle*{1}}\put(35.18,29){\circle*{1}}
\put(35.42,31){\circle*{1}}\put(35.77,33){\circle*{1}}
\put(35.05,23){\circle*{1}}\put(35.18,21){\circle*{1}}
\put(37,13){\circle*{1}} 
\qbezier(38.2,39)(32.79,32.21)(40,25)\qbezier(43.8,11)(47.21,17.79)(40,25)
\put(41.2,26.2){\circle*{1}}\put(42.4,27.6){\circle*{1}}
\put(43.5,29.5){\circle*{1}}\put(44.3,31.4){\circle*{1}}
\put(45,33.8){\circle*{1}}
\put(38.8,23.8){\circle*{1}}\put(37.6,22.4){\circle*{1}}
\put(36.5,20.5){\circle*{1}}
\end{picture}
\end{center}
\end{figure}
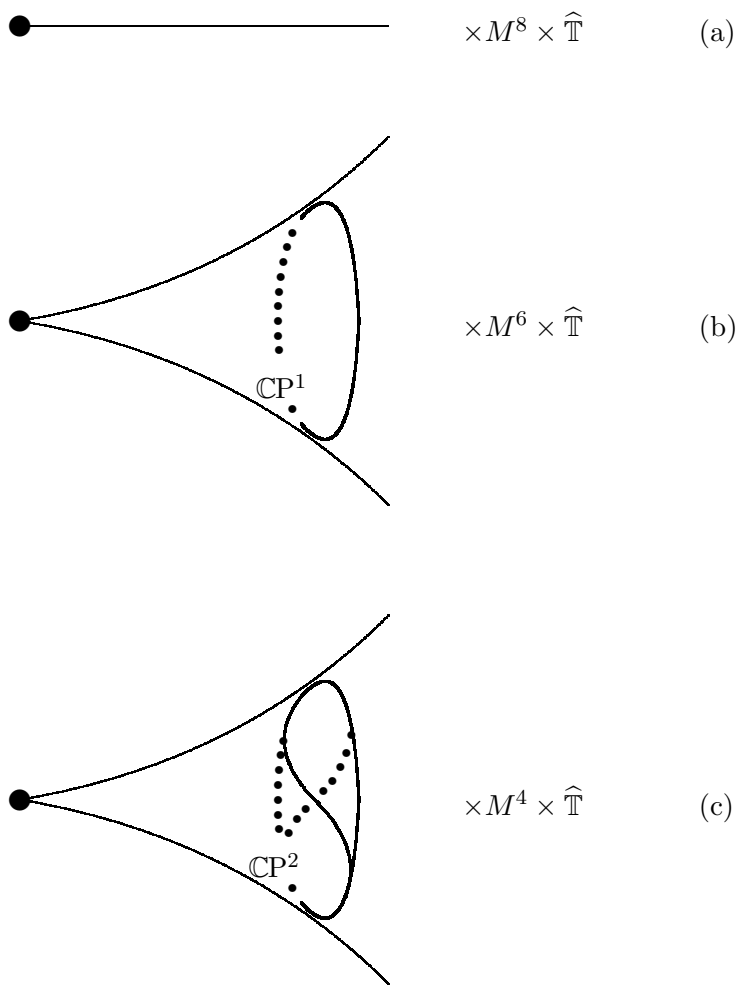

\newpage

\noindent{\bf Appendix}\\

\noindent The appropriate K-theory for countably compactly generated spaces and
$\sC$-algebras is representable K-theory, denoted by $\RK$, cf.~\cite{P3}.
The RK-theory agrees with the usual K-theory of compact spaces and
$C^*$-algebras for a compact space or a $C^*$-algebra.
Many of the nice properties that K-theory satisfies generalise to RK-theory:

\begin{enumerate}
\item\label{homotopyinv}
The $\RK$-theory is homotopy invariant and satisfies Bott periodicity, hence
it is $\ZZ_2 $-graded.
\item\label{K}
If $A$ is a $C^*$-algebra, then there is a natural isomorphism
$\RK_i(A)\cong\K_i(A)$. 
\item
There is a natural isomorphism $\RK_i(A\otimes \cpt)\cong\RK_i(A)$, where
$\cpt$ denotes the algebra of compact operators on a separable Hilbert space.
\end{enumerate}

Let $\CT(X,H)$ denote the continuous trace $C^*$-algebra on $X$ consisting of
the sections of an algebra bundle of compact operators over $X$, with
Dixmier-Douady class $H$.
Then the twisted $\K$-theory is by definition \cite{Ros}
$$   \K^\bullet(X,H)=\K_\bullet(\CT(X,H)).   $$ 
Now $\CT(X\times E\TT,p_1^*(H))=\varprojlim_n\CT(X\times S^{2n+1},p_1^*(H))$ is
a $\sC$-algebra and the twisted $\RK$-theory is
$$  \RK^\bullet(X\times E\TT,p_1^*(H))=\RK_\bullet(\CT(X\times E\TT,p_1^*(H))). $$
Since $\CT(X\times E\TT,p_1^*(H))=\CT(X,H)\otimes C(E\TT)$ and $C(E\TT)$ is
contractible, by the homotopy invariance of representable K-theory in property
(\ref{homotopyinv}) above,
$$ \RK^\bullet(X\times E\TT,p_1^*(H))\cong\RK^\bullet(X,H) $$
and by property (\ref{K}) above, we see that
$$ \RK^\bullet(X,H)\cong\K^\bullet(X,H). $$
Combining the observations above, we see that
$$ \K^\bullet(X,H)\cong\RK^\bullet(X\times E\TT,p_1^*(H)). $$
By the Connes-Thom isomorphism for $\sC$-algebras \cite{PS}, 
$$      \RK^\bullet(X\times E\TT,p_1^*(H))\cong
     \RK_{\bullet+1}(\CT(X\times E\TT,p_1^*(H))\rtimes\RR).    $$
Observe that one has the commutative diagram of finite dimensional
approximations of \eqref{correspondenceb}, 
\begin{equation}\label{correspondencec}
\qquad\xymatrix @=4pc @ur {X\times S^{2n+1}\ar[d]_{\pi} & 
\widehat X\times S^{2n+1}\ar[d]_{\hat p}\ar[l]^{p} \\
X_\TT(n). & \widehat X_\TT(n)\ar[l]^{\hat \pi}}.
\end{equation}
By \cite{RR88}, one has 
$$  \CT(X\times S^{2n+1},p_1^*(H))\rtimes\RR\cong
    \CT(\widehat X_\TT(n),\widehat H(n)),      $$
and since
\begin{align*} 
\CT(\widehat X_\TT,\widehat H)
    &=\varprojlim_n\CT(\widehat X_\TT(n),\widehat H(n)),\\
    \CT(X\times E\TT,p_1^*(H))\rtimes\RR
    &=\varprojlim_n \CT(X\times S^{2n+1},p_1^*(H))\rtimes\RR,     
 \end{align*}
we deduce that
$$ \CT(X\times E\TT,p_1^*(H))\rtimes\RR\cong\CT(\widehat X_\TT,\widehat H), $$
and hence that
\beq\label{T-dual1}
\qquad\K^\bullet(X,H)\cong\RK^{\bullet+1}(\widehat X_\TT,\widehat H).
\eeq
Next, the $I(\TT)$-adic completion $\K_\TT^\bullet(\widehat X, \widehat H)^\sim$
of twisted equivariant K-theory (where $I(\TT)$ denotes the augmentation ideal)
is isomorphic to, cf.~\cite{PAS},
\beq\label{T-dual2}
\qquad\K_\TT^\bullet(\widehat X, \widehat H)^\sim
\cong\RK^\bullet(\widehat X_\TT,\widehat H).
\eeq
Therefore using \eqref{T-dual1} and \eqref{T-dual2} we get
\beq\label{T-dual}
\qquad\K^\bullet(X,H)\cong\K_\TT^\bullet(\widehat X,\widehat H)^\sim,
\eeq
which is interpreted as T-duality isomorphism of charges of RR-fields in
K-theory.

\end{document}